\documentstyle[prd,aps,preprint,tighten,epsfig]{revtex}

\begin{document}

\draft

\title{Towards Determination of the Initial Flavor Composition of
Ultrahigh-energy Neutrino Fluxes with Neutrino Telescopes}
\author{{\bf Zhi-zhong Xing}
\thanks{E-mail: xingzz@mail.ihep.ac.cn} ~ and ~ {\bf Shun Zhou}
\thanks{E-mail: zhoush@mail.ihep.ac.cn}}
\address{CCAST (World Laboratory), P.O. Box 8730, Beijing 100080,
China \\
and Institute of High Energy Physics, Chinese Academy of Sciences, \\
P.O. Box 918, Beijing 100049, China}

\maketitle

\begin{abstract}
We propose a simple but useful parametrization of the flavor
composition of ultrahigh-energy neutrino fluxes produced from
distant astrophysical sources: $\phi^{}_e : \phi^{}_\mu :
\phi^{}_\tau = \sin^2 \xi \cos^2 \zeta : \cos^2 \xi \cos^2 \zeta :
\sin^2 \zeta$. We show that it is possible to determine or
constrain $\xi$ and $\zeta$ by observing two independent neutrino
flux ratios at the second-generation neutrino telescopes, provided
three neutrino mixing angles and the Dirac CP-violating phase have
been well measured in neutrino oscillations. Any deviation of
$\zeta$ from zero will signify the existence of cosmic
$\nu^{}_\tau$ and $\overline{\nu}^{}_\tau$ neutrinos at the
source, and an accurate value of $\xi$ can be used to test both
the conventional mechanism and the postulated scenarios for cosmic
neutrino production.
\end{abstract}


\newpage

\section{Introduction}

High-energy neutrino telescopes are going to open a new window on
the Universe, since they can be used to probe and characterize
very distant astrophysical sources \cite{Review}. The promising
IceCube neutrino telescope \cite{Halzen}, which has a
kilometer-scale detector, is now under construction. If the
relative fluxes of ultrahigh-energy $\nu^{}_e$
($\overline{\nu}^{}_e$), $\nu^{}_\mu$ ($\overline{\nu}^{}_\mu$)
and $\nu^{}_\tau$ ($\overline{\nu}^{}_\tau$) neutrinos are
successfully measured at IceCube and other neutrino telescopes, it
will be possible to diagnose the relevant cosmic accelerators
(e.g., their locations and characteristics) and examine the
properties of neutrinos themselves (e.g., neutrino mixing and
leptonic CP violation).

Indeed, robust evidence for neutrino masses and lepton flavor
mixing has been achieved from the recent solar \cite{SNO},
atmospheric \cite{SK}, reactor \cite{KM} and accelerator
\cite{K2K} neutrino oscillation experiments. Due to neutrino
oscillations, the neutrino fluxes observed at the detector
$\Phi^{\rm D} = \{\phi^{\rm D}_e, \phi^{\rm D}_\mu, \phi^{\rm
D}_\tau \}$ are in general different from the source fluxes $\Phi
= \{\phi^{}_e, \phi^{}_\mu, \phi^{}_\tau \}$. Note that our
notation is $\phi^{(\rm D)}_\alpha \equiv \phi^{(\rm
D)}_{\nu^{}_\alpha} + \phi^{(\rm D)}_{\overline{\nu}^{}_\alpha}$
(for $\alpha = e, \mu, \tau$), where $\phi^{(\rm
D)}_{\nu^{}_\alpha}$ and $\phi^{(\rm
D)}_{\overline{\nu}^{}_\alpha}$ denote the
$\nu^{}_\alpha$-neutrino and
$\overline{\nu}^{}_\alpha$-antineutrino fluxes, respectively. The
relation between $\phi^{}_{\nu^{}_\alpha}$ (or
$\phi^{}_{\overline{\nu}^{}_\alpha}$) and $\phi^{\rm
D}_{\nu^{}_\beta}$ (or $\phi^{\rm D}_{\overline{\nu}^{}_\beta}$)
is given by
\begin{eqnarray}
\phi^{\rm D}_{\nu^{}_\beta} & = & \sum_\alpha \left (
\phi^{}_{\nu^{}_\alpha} P^{}_{\alpha \beta} \right ) \; ,
\nonumber \\
\phi^{\rm D}_{\overline{\nu}^{}_\beta} & = & \sum_\alpha \left (
\phi^{}_{\overline{\nu}^{}_\alpha} \bar{P}^{}_{\alpha \beta}
\right ) \; ,
\end{eqnarray}
in which $P^{}_{\alpha \beta}$ and $\bar{P}^{}_{\alpha \beta}$
stand respectively for the oscillation probabilities $P
(\nu^{}_\alpha \rightarrow \nu^{}_\beta)$ and $P
(\overline{\nu}^{}_\alpha \rightarrow \overline{\nu}^{}_\beta)$.
As the Galactic distances far exceed the observed neutrino
oscillation lengths, $P^{}_{\alpha \beta}$ and $\bar{P}^{}_{\alpha
\beta}$ are actually averaged over many oscillations and take a
very simple form:
\begin{equation}
P^{}_{\alpha \beta} \; = \; \bar{P}^{}_{\alpha \beta} \; = \;
\sum^3_{i=1} |V^{}_{\alpha i}|^2 |V^{}_{\beta i}|^2 \; ,
\end{equation}
where $V^{}_{\alpha i}$ and $V^{}_{\beta i}$ (for $\alpha, \beta =
e, \mu, \tau$ and $i = 1, 2, 3$) are just the elements of the
$3\times 3$ neutrino mixing matrix $V$. Eqs. (1) and (2) lead us
to a straightforward relation between $\phi^{}_\alpha$ and
$\phi^{\rm D}_\beta$:
\begin{equation}
\phi^{\rm D}_\beta \; = \; \sum_\alpha \left ( \phi^{}_\alpha
P^{}_{\alpha \beta} \right ) \; . ~
\end{equation}
This relation indicates that the observation of $\Phi^{\rm D}$ at
a neutrino telescope can {\it at least} help
\footnote{One may also use neutrino telescopes to test the
stability of neutrinos \cite{Beacom,Quigg}, possible violation of
CPT symmetry \cite{Quigg} and other exotic scenarios of particle
physics and cosmology.}
\begin{itemize}
\item      to determine or constrain the flavor composition of
cosmic neutrino fluxes \cite{Quigg,Beacom2}, if three neutrino
mixing angles and the Dirac CP-violating phase hidden in
$P^{}_{\alpha \beta}$ have been measured to a good degree of
accuracy (e.g., a precision of $10\%$ or smaller relative error
bars \cite{Winter});
\end{itemize}
or
\begin{itemize}
\item      to determine or constrain one or two of three neutrino
mixing angles and the Dirac CP-violating phase
\cite{Gupta,Serpico}, provided the production mechanism of
ultrahigh-energy neutrinos at an astrophysical source (e.g., the
conventional source to be mentioned below) is really understood.
\end{itemize}
Hence neutrino telescopes will serve as a very useful tool to
probe both the high-energy astrophysical processes and the
intrinsic properties of massive neutrinos.

This paper aims at a determination of the flavor composition of
cosmic neutrino fluxes at the source with the help of neutrino
telescopes. Different from the previous works (see, e.g., Refs.
\cite{Beacom,Quigg,Beacom2,Gupta,Serpico}), our present study
starts from a generic parametrization of the initial neutrino
fluxes:
\begin{equation}
\left \{ \phi^{}_e ~ , ~\phi^{}_\mu ~ , ~\phi^{}_\tau \right \} \;
= \; \left \{ \sin^2 \xi \cos^2 \zeta ~ , ~\cos^2 \xi \cos^2 \zeta
~ , ~\sin^2 \zeta \right \} \phi^{}_0 \; ,
\end{equation}
where $\xi \in [0, \pi/2]$ and $\zeta \in [0, \pi/2]$, and
$\phi^{}_0$ denotes the total flux (i.e., the sum of three
neutrino fluxes). Provided ultrahigh-energy neutrinos are produced
by certain astrophysical sources (e.g., Active Galactic Nuclei or
AGN) via the decay of pions created from $pp$ and $p\gamma$
collisions, for instance, their flavor content is expected to be
\begin{equation}
\left \{\phi^{}_e : \phi^{}_\mu : \phi^{}_\tau \right \} \; = \;
\left \{ \frac{1}{3} : \frac{2}{3} : 0 \right \} \; .
\end{equation}
This ``standard" neutrino flux ratio corresponds to $\zeta = 0$
and $\tan\xi = 1/\sqrt{2}$ (or equivalently $\xi \approx
35.3^\circ$) in our parametrization. It turns out that any small
departure of $\zeta$ from zero will measure the existence of
cosmic $\nu^{}_\tau$ and $\overline{\nu}^{}_\tau$ neutrinos, which
might come from the decays of $D^{}_s$ and $B\overline{B}$ mesons
produced at the source \cite{Pakvasa}. On the other hand, any
small deviation of $\tan^2\xi$ from $1/2$ will imply that the
conventional mechanism for ultrahigh-energy neutrino production
from the AGN has to be modified. Similar arguments can be put
forward for the neutrino fluxes from some other astrophysical
sources, such as the postulated neutron beam source \cite{Neutron}
with
\begin{equation}
\left \{\phi^{}_e : \phi^{}_\mu : \phi^{}_\tau \right \} \; = \;
\left \{ 1 : 0 : 0 \right \} \;
\end{equation}
(or equivalently $\{ \xi, \zeta \} = \{ \pi/2, 0 \}$) and the
possible muon-damped source \cite{Muon} with
\begin{equation}
\left \{\phi^{}_e : \phi^{}_\mu : \phi^{}_\tau \right \} \; = \;
\left \{ 0 : 1 : 0 \right \} \;
\end{equation}
(or equivalently $\{ \xi, \zeta \} = \{ 0, 0 \}$). Therefore, we
are well motivated to investigate how the {\it true} values of
$\xi$ and $\zeta$ for a specific astrophysical source can be
determined or constrained by use of the second-generation neutrino
telescopes and with the help of more precise data from the
upcoming long-baseline neutrino oscillation experiments. This goal
is indeed reachable, as we shall explicitly demonstrate in the
remaining part of this paper.

The remaining part of this paper is organized as follows. In
section II, we derive the analytical relations between the
neutrino flavor parameters ($\xi$ and $\zeta$) at an astrophysical
source and the typical observables of neutrino fluxes at a
terrestrial detector. Section III is devoted to a detailed
numerical analysis of the dependence of those observables on $\xi$
and $\zeta$. A brief summary of our main results is given in
section IV.

\section{Observables}

Because of neutrino oscillations and the $\nu^{}_\tau$
``regeneration" in the Earth \cite{Halzen}, it is especially
important to detect all three flavors of the cosmic neutrinos at a
neutrino telescope. The sum of $\phi^{\rm D}_e$, $\phi^{\rm
D}_\mu$ and $\phi^{\rm D}_\tau$ is equal to that of $\phi^{}_e$,
$\phi^{}_\mu$ and $\phi^{}_\tau$,
\begin{equation}
\phi^{}_0 \; \equiv \; \phi^{}_e + \phi^{}_\mu + \phi^{}_\tau =
\phi^{\rm D}_e + \phi^{\rm D}_\mu + \phi^{\rm D}_\tau \; ,
\end{equation}
as one may easily see from Eqs. (2) and (3). A measurement of
$\phi^{}_0$ may involve large systematic uncertainties, but the
latter can be largely cancelled out in the ratio of two neutrino
fluxes. Therefore, let us follow Ref. \cite{Serpico} to define
\begin{equation}
\left \{ R^{}_e ~ , ~ R^{}_\mu ~ , ~R^{}_\tau \right \} \; \equiv
\; \left \{ \frac{\phi^{\rm D}_e}{\phi^{\rm D}_\mu + \phi^{\rm
D}_\tau} ~ , ~\frac{\phi^{\rm D}_\mu}{\phi^{\rm D}_\tau +
\phi^{\rm D}_e} ~ , ~\frac{\phi^{\rm D}_\tau}{\phi^{\rm D}_e +
\phi^{\rm D}_\mu} \right \}  \;
\end{equation}
as our {\it working} observables. We remark that these ratios are
largely free from the systematic uncertainties associated with the
measurements of $\phi^{\rm D}_e$, $\phi^{\rm D}_\mu$ and
$\phi^{\rm D}_\tau$. In particular, it is relatively easy to
extract $R^{}_\mu$ from the ratio of muon tracks to showers at
IceCube \cite{IC}, even if those electron and tau events may not
well be disentangled. Since $R^{}_e$, $R^{}_\mu$ and $R^{}_\tau$
satisfy
\begin{equation}
\frac{R^{}_e}{1+R^{}_e} + \frac{R^{}_\mu}{1+R^{}_\mu} +
\frac{R^{}_\tau}{1+R^{}_\tau} \; = \; 1 \; ,
\end{equation}
only two of them are independent.

Note that Eqs. (6) and (7) represent two peculiar (non-standard)
scenarios of cosmic neutrino production. In principle, one may
also assume an exotic astrophysical source which only produces
$\nu^{}_\tau$ and $\overline{\nu}^{}_\tau$ neutrinos; i.e.,
$\{\phi^{}_e : \phi^{}_\mu : \phi^{}_\tau \} = \{ 0 : 0 : 1 \}$ or
equivalently $\zeta = \pi/2$ with unspecified $\xi$ in our
parametrization. The expression of $R^{}_\alpha$ (for $\alpha = e,
\mu, \tau$) can then be simplified in such special cases:
\begin{equation}
R^{}_\alpha \; =\; \left \{ \begin{array}{l} \displaystyle
\frac{P^{}_{e\alpha}}{1 - P^{}_{e\alpha}} \; , ~~~~ {\rm for} ~ \{
\xi, \zeta \} = \{ \pi/2, 0 \} \; , \\
\displaystyle \frac{P^{}_{\mu\alpha}}{1 - P^{}_{\mu\alpha}} \; ,
~~~~ {\rm for} ~ \{
\xi, \zeta \} = \{ 0, 0 \} \; , \\
\displaystyle \frac{P^{}_{\tau\alpha}}{1 - P^{}_{\tau\alpha}} \; ,
~~~~ {\rm for} ~ \{ \xi, \zeta \} = \{ *, \pi/2 \} \; .
\end{array}
\right .
\end{equation}
Even if the third case is completely unrealistic, it could serve
as an example to illustrate the salient feature of $R^{}_\alpha$
defined above.

Without loss of generality, we choose $R^{}_e$ and $R^{}_\mu$ as
two typical observables and derive their explicit relations with
$\xi$ and $\zeta$. By using Eqs. (4), (8) and (9), we obtain
\begin{eqnarray}
R^{}_e & = & \frac{P^{}_{e e} \sin^2\xi + P^{}_{\mu e} \cos^2\xi +
P^{}_{\tau e} \tan^2\zeta}{\sec^2\zeta - \left[ P^{}_{e e} \sin^2\xi
+ P^{}_{\mu e} \cos^2\xi + P^{}_{\tau e} \tan^2\zeta \right]} \; ,
\nonumber \\
R^{}_\mu & = & \frac{P^{}_{e \mu} \sin^2\xi + P^{}_{\mu \mu}
\cos^2\xi + P^{}_{\tau \mu} \tan^2\zeta}{\sec^2\zeta - \left[
P^{}_{e \mu} \sin^2\xi + P^{}_{\mu \mu} \cos^2\xi + P^{}_{\tau \mu}
\tan^2\zeta \right]} \; .
\end{eqnarray}
The source flavor parameters $\xi$ and $\zeta$ can then be figured
out in terms of $R^{}_e$ and $R^{}_\mu$:
\begin{eqnarray}
\sin^2\xi & = & \frac{r^{}_e \left ( P^{}_{\tau \mu} - P^{}_{\mu
\mu} \right ) - r^{}_\mu \left ( P^{}_{\tau e} - P^{}_{\mu e}
\right ) + \left ( P^{}_{\mu \mu} P^{}_{\tau e} - P^{}_{\mu e}
P^{}_{\tau \mu} \right )}{\left ( r^{}_e - P^{}_{\tau e} \right )
\left ( P^{}_{e \mu} - P^{}_{\mu \mu} \right ) - \left ( r^{}_\mu
- P^{}_{\tau \mu} \right ) \left ( P^{}_{e e} - P^{}_{\mu e}
\right )} \; ,
\nonumber \\
\tan^2\zeta & = & \frac{r^{}_e \left ( P^{}_{\mu \mu} - P^{}_{e
\mu} \right ) - r^{}_\mu \left ( P^{}_{\mu e} - P^{}_{e e} \right
) + \left ( P^{}_{e \mu} P^{}_{\mu e} - P^{}_{e e} P^{}_{\mu \mu}
\right )} {\left ( r^{}_e - P^{}_{\tau e} \right ) \left ( P^{}_{e
\mu} - P^{}_{\mu \mu} \right ) - \left ( r^{}_\mu - P^{}_{\tau
\mu} \right ) \left ( P^{}_{e e} - P^{}_{\mu e} \right )} \; ,
\end{eqnarray}
where the notations $r^{}_e \equiv R^{}_e/(1 + R^{}_e)$ and
$r^{}_\mu \equiv R^{}_\mu/(1 + R^{}_\mu)$ have been used to
simplify the expressions. Indeed, $r^{}_e = \phi^{\rm
D}_e/\phi^{}_0$ and $r^{}_\mu = \phi^{\rm D}_\mu/\phi^{}_0$ hold.
One may in principle choose either $(R^{}_e, R^{}_\mu)$ or
$(r^{}_e, r^{}_\mu)$ as a set of working observables to inversely
determine $\xi$ and $\zeta$. The first set has been chosen by a
few authors (see, e.g., Refs. \cite{Winter,Serpico}) and will also
be adopted in this paper. Note that the averaged neutrino
oscillation probabilities $P^{}_{\alpha \beta}$ depend on three
mixing angles $(\theta^{}_{12}, \theta^{}_{23}, \theta^{}_{13})$
and the Dirac CP-violating phase $(\delta)$ in the standard
parametrization of the $3\times 3$ neutrino mixing matrix $V$
\cite{PDG}. Taking account of $\theta^{}_{13} < 10^\circ$
\cite{CHOOZ} but $\theta^{}_{12} \approx 34^\circ$ and
$\theta^{}_{23} \approx 45^\circ$ \cite{Vissani}, we express
$P^{}_{\alpha \beta}$ as the first-order expansion of the small
parameter $\sin\theta^{}_{13}$:
\begin{eqnarray}
P^{}_{e e} &=& 1 - \frac{1}{2}\sin^2 2\theta^{}_{12} \; ,
\nonumber \\
P^{}_{e \mu} &=& \frac{1}{2} \sin^2 2\theta^{}_{12} \cos^2
\theta^{}_{23} + \frac{1}{4} \sin 4\theta^{}_{12} \sin
2\theta^{}_{23} \sin \theta^{}_{13} \cos \delta \; ,
\nonumber \\
P^{}_{e \tau} &=& \frac{1}{2} \sin^2 2\theta^{}_{12} \sin^2
\theta^{}_{23} - \frac{1}{4} \sin 4\theta^{}_{12} \sin
2\theta^{}_{23} \sin \theta^{}_{13} \cos \delta \; ,
\nonumber \\
P^{}_{\mu \mu} &=& 1 - \frac{1}{2} \sin^2 2\theta^{}_{23}
-\frac{1}{2} \sin^2 2\theta^{}_{12} \cos^4 \theta^{}_{23} -
\frac{1}{2} \sin 4 \theta^{}_{12} \sin 2\theta^{}_{23} \cos^2
\theta^{}_{23} \sin \theta^{}_{13} \cos \delta \; ,
\nonumber \\
P^{}_{\mu \tau} &=& \frac{1}{2} \sin^2 2\theta^{}_{23} -
\frac{1}{8} \sin^2 2\theta^{}_{12} \sin^2 2\theta^{}_{23} +
\frac{1}{8} \sin 4\theta^{}_{12} \sin 4\theta^{}_{23} \sin
\theta^{}_{13} \cos \delta \; ,
\nonumber \\
P^{}_{\tau \tau} &=& 1 - \frac{1}{2} \sin^2 2\theta^{}_{23}
-\frac{1}{2} \sin^2 2\theta^{}_{12} \sin^4 \theta^{}_{23} +
\frac{1}{2} \sin 4 \theta^{}_{12} \sin 2\theta^{}_{23} \sin^2
\theta^{}_{23} \sin \theta^{}_{13} \cos \delta \; ,
\end{eqnarray}
in which the higher-order terms, such as ${\cal
O}(\sin^2\theta^{}_{13}) < 3\%$, have been safely neglected. One can
see that the sensitivity of $P^{}_{\alpha \beta}$ to $\delta$ is
suppressed due to the smallness of $\sin\theta^{}_{13}$. Hence the
dependence of $R^{}_{\alpha}$ on $\delta$ is expected to be
insignificant. In addition, it is hard to distinguish between the
cases of $\theta^{}_{13} = 0^\circ$ and $\delta^{} = 90^\circ$,
because it is the product of $\sin\theta^{}_{13}$ and $\cos\delta$
that appears in the analytical approximations of $P^{}_{\alpha
\beta}$.

A global analysis of current neutrino oscillation data
\cite{Vissani} yields
\begin{eqnarray}
30^\circ  < & \theta^{}_{12} & < \; 38^\circ \; ,
\nonumber \\
36^\circ < & \theta^{}_{23} &  < \; 54^\circ \; ,
\nonumber \\
0^\circ \leq & \theta^{}_{13} &  < \; 10^\circ \; ,
\end{eqnarray}
at the $99\%$ confidence level. The best-fit values of three
neutrino mixing angles are $\theta^{}_{12} \approx 34^\circ$,
$\theta^{}_{23} \approx 45^\circ$ and $\theta^{}_{13} \approx
0^\circ$ \cite{Vissani}, but the CP-violating phase $\delta$ is
entirely unrestricted. Although the present experimental data
remain unsatisfactory, they can be used to constrain the
correlation between the parameters $(\xi, \zeta)$ and the
observables $(R^{}_\alpha, R^{}_\beta)$. We shall illustrate our
analytical results by taking a few typical numerical examples in
the subsequent section.

\section{Illustration}

First of all, let us follow a rather conservative strategy to scan
the reasonable ranges of $(\theta^{}_{12}, \theta^{}_{23},
\theta^{}_{13}, \delta)$ and $(\xi, \zeta)$ so as to examine the
sensitivities of $(R^{}_e, R^{}_\mu, R^{}_\tau)$ to these six
parameters. We take $\delta \in [0^\circ, 180^\circ]$ in addition
to the generous intervals of three mixing angles given in Eq.
(15), and allow $\xi$ to vary in the region $\xi \in [0^\circ,
90^\circ]$. As the amount of $\nu^{}_\tau$ and
$\overline{\nu}^{}_\tau$ neutrinos produced at those realistic
astrophysical sources is expected to be very small or even
vanishing, we restrict the parameter $\zeta$ to a very narrow
domain $\zeta \in [0^\circ, 18^\circ]$, which corresponds to
$\sin^2 \zeta \lesssim 0.1$ \cite{Pakvasa}. It should be noted
that we do the numerical computation by using the exact
expressions of $P^{}_{\alpha\beta}$ instead of Eq. (14). Our
results are shown in FIG. 1. Two comments are in order:
\begin{enumerate}
\item     The source parameter $\xi$ is quite sensitive to the
values of the neutrino flux ratios $R^{}_\alpha$ (for $\alpha = e,
\mu, \tau$). Even if three neutrino mixing angles involve a lot of
uncertainties and the Dirac CP-violating phase is entirely unknown,
a combined measurement of $(R^{}_e, R^{}_\mu)$ or $(R^{}_e,
R^{}_\tau)$ can constrain the value of $\xi$ to an acceptable degree
of accuracy. This encouraging observation assures that the
second-generation neutrino telescopes can really be used to probe
the initial flavor composition of ultrahigh neutrino fluxes. Note
that the standard pion-decay source $\{\phi^{}_e : \phi^{}_\mu :
\phi^{}_\tau \} = \{ 1/3 : 2/3 : 0 \}$ will produce $\{\phi^{\rm
D}_e : \phi^{\rm D}_\mu : \phi^{\rm D}_\tau \} \simeq \{ 1/3 : 1/3 :
1/3 \}$ or equivalently $R^{}_e \approx R^{}_\mu \approx R^{}_\tau
\approx 0.5$ at the detector, as shown in FIG. 1, if $\theta^{}_{23}
\approx 45^\circ$ is fixed and $\theta^{}_{13} < 10^\circ$ is taken.
Such an expectation cannot be true, however, when $\xi$ deviates
from its given value $\xi = 35.3^\circ$ and (or) when
$\theta^{}_{23}$ departs from its best-fit value $\theta^{}_{23} =
45^\circ$. More precise neutrino oscillation data will greatly help
to narrow down the $(R^{}_\alpha, \xi)$ parameter space.

\item     In contrast, the source parameter $\zeta$ seems to be
insensitive to $R^{}_\alpha$ (for $\alpha = e, \mu, \tau$). The
reason for this insensitivity is two-fold: (a) the values of
$\zeta$ have been restricted to a very narrow range ($0^\circ \leq
\zeta \leq 18^\circ$); and (b) the numerical uncertainties of
three neutrino mixing angles and the Dirac CP-violating phase are
too large. Provided $\theta^{}_{12}$, $\theta^{}_{23}$,
$\theta^{}_{13}$ and $\delta$ are all measured to a high degree of
accuracy in the near future, it will be possible to find out the
definite dependence of $R^{}_\alpha$ on $\zeta$ for a given value
of $\xi$.
\end{enumerate}
To be more explicit, we are going to consider three typical
scenarios of cosmic neutrino fluxes and illustrate the
sensitivities of $R^{}_e$, $R^{}_\mu$ and $R^{}_\tau$ to the
source parameters $(\xi, \zeta)$ and to the unknown neutrino
mixing parameters $(\theta^{}_{13}, \delta)$.

We argue that the simple flavor content of ultrahigh-energy
neutrino fluxes from the standard pion-decay source could somehow
be contaminated for certain reasons: e.g., a small amount of
$\nu^{}_e$, $\nu^{}_\mu$ and $\nu^{}_\tau$ and their antiparticles
might come from the decays of heavier hadrons produced by $pp$ and
$p\gamma$ collisions \cite{Pakvasa}. Similar arguments can also be
made for the postulated neutron beam source and the possible
muon-damped source, as our present knowledge about the mechanism
of cosmic neutrino production remains very poor. Following a
phenomenological approach, we slightly modify the scenarios listed
in Eqs. (5), (6) and (7) by allowing the relevant $\xi$ and
$\zeta$ parameters to fluctuate around their given values. Namely,
we consider
\begin{itemize}
\item     {\bf Scenario A:} $30^\circ \leq \xi \leq 40^\circ$ and
$0^\circ \leq \zeta \leq 18^\circ$, serving as a modified version
of the standard pion-decay source (originally, $\xi = 35.3^\circ$
and $\zeta = 0^\circ$);

\item     {\bf Scenario B:} $80^\circ \leq \xi \leq 90^\circ$ and
$0^\circ \leq \zeta \leq 18^\circ$, serving as a modified version
of the postulated neutron beam source (originally, $\xi =
90^\circ$ and $\zeta = 0^\circ$);

\item     {\bf Scenario C:} $0^\circ \leq \xi \leq 10^\circ$ and
$0^\circ \leq \zeta \leq 18^\circ$, serving as a modified version
of the possible muon-damped source (originally, $\xi = 0^\circ$
and $\zeta = 0^\circ$).
\end{itemize}
For simplicity, we fix $\theta^{}_{12} = 34^\circ$ and
$\theta^{}_{23} = 45^\circ$ in our numerical analysis. We take
four typical inputs for the unknown parameters $\theta^{}_{13}$
and $\delta$: (a) $\theta^{}_{13} = 0^\circ$ (in this case,
$\delta$ is not well-defined and has no physical significance);
(b) $\theta^{}_{13} = 5^\circ$ and $\delta = 0^\circ$; (c)
$\theta^{}_{13} = 5^\circ$ and $\delta = 90^\circ$; and (d)
$\theta^{}_{13} = 5^\circ$ and $\delta = 180^\circ$. Our numerical
results for the sensitivities of $R^{}_e$, $R^{}_\mu$ and
$R^{}_\tau$ to $\xi$ and $\zeta$ in scenarios A, B and C are shown
in FIGs. 2, 3 and 4, respectively. Some discussions are in order.
\begin{enumerate}
\item   \underline{Scenario A in FIG. 2}. The neutrino flux ratios
$R^{}_e$, $R^{}_\mu$ and $R^{}_\tau$ are all sensitive to the
small deviation of $\xi$ from its standard value $\xi =
35.3^\circ$. In contrast, the changes of three observables are
very small when $\zeta$ varies from $0^\circ$ to $18^\circ$. This
insensitivity is understandable: about half of the initial
$\nu^{}_\mu$ and $\overline{\nu}^{}_\mu$ neutrinos oscillate into
$\nu^{}_\tau$ and $\overline{\nu}^{}_\tau$ neutrinos, whose amount
dominates over the survival amount of initial $\nu^{}_\tau$ and
$\overline{\nu}^{}_\tau$ neutrinos at the detector. In other
words, $\phi^{\rm D}_\tau \gg \phi^{}_\tau$ and $\phi^{\rm D}_e
\sim \phi^{\rm D}_\mu \sim \phi^{\rm D}_\tau$ hold, implying that
$R^{}_\alpha$ must be insensitive to small $\phi^{}_\tau$ or
equivalently to small $\zeta$. It is therefore difficult to pin
down the value of $\zeta$ from this kind of astrophysical sources.
As pointed out in the last section, the result of $R^{}_\alpha$ in
the $\theta^{}_{13} = 0^\circ$ case (solid curves) is almost
indistinguishable from that in the $\delta = 90^\circ$ case
(dotted curves). The sensitivity of $R^{}_\alpha$ to $\delta$ is
insignificant but distinguishable, if the value of
$\theta^{}_{13}$ is about $5^\circ$ or larger.

\item   \underline{Scenario B in FIG. 3}. The neutrino flux ratio
$R^{}_e$ is sensitive to the small departures of $\xi$ and $\zeta$
from their given values $\xi = 90^\circ$ and $\zeta = 0^\circ$.
This salient feature can be understood as follows. Since
$\nu^{}_\mu$ (or $\overline{\nu}^{}_\mu$) and $\nu^{}_\tau$ (or
$\overline{\nu}^{}_\tau$) neutrinos at the detector mainly come
from the initial $\nu^{}_e$ (or $\overline{\nu}^{}_e$) neutrinos
via the oscillation, the sum of $\phi^{\rm D}_\mu$ and $\phi^{\rm
D}_\tau$ is expected to be smaller than or comparable with the
survival $\nu^{}_e$ (or $\overline{\nu}^{}_e$) flux $\phi^{\rm
D}_e$. The roles of $\phi^{}_e$ and $\phi^{}_\tau$ are important
in $\phi^{\rm D}_e$ and $\phi^{\rm D}_\mu + \phi^{\rm D}_\tau$,
respectively. It turns out that $R^{}_e$ depends, in a relatively
sensitive way, on $\xi$ through $\phi^{\rm D}_e$ in its numerator
and on $\zeta$ through $\phi^{\rm D}_\mu + \phi^{\rm D}_\tau$ in
its denominator. Note also that the nearly degenerate results of
$R^{}_e$ for $\theta^{}_{13} = 5^\circ$ and $\delta = (0^\circ,
90^\circ, 180^\circ)$ are primarily attributed to the fact that
$P^{}_{ee}$ in the numerator of $R^{}_e$ is actually independent
of $\delta$. On the other hand, the discrepancy between $R^{}_e
(\theta^{}_{13}=0^\circ)$ and $R^{}_e (\theta^{}_{13}=5^\circ)$ in
FIG. 3 results from the ${\cal O}(\sin^2\theta^{}_{13})$ terms of
$P^{}_{\alpha \beta}$, which have been neglected in Eq. (14). In
comparison with $R^{}_e$, the neutrino flux ratios $R^{}_\mu$ and
$R^{}_\tau$ are not so sensitive to the small changes of $\xi$ and
$\zeta$. But the measurement of $R^{}_\mu$ and $R^{}_\tau$ is as
important as that of $R^{}_e$, in order to determine the flavor
composition of ultrahigh-energy neutrino fluxes at such an
astrophysical source.

\item   \underline{Scenario C in FIG. 4}. In this case, in which
$\phi^{\rm D}_e$ and $\phi^{\rm D}_\tau$ mainly come from the
initial $\phi^{}_\mu$ via $\nu^{}_\mu \rightarrow \nu^{}_e$ and
$\nu^{}_\mu \rightarrow \nu^{}_\tau$ oscillations, one can
similarly understand the numerical behaviors of $R^{}_e$,
$R^{}_\mu$ and $R^{}_\tau$ changing with the small deviation of
$\xi$ from its given value $\xi = 0^\circ$. To explain why three
observables are almost independent of the fluctuation of $\zeta$,
we take $\theta^{}_{23} = 45^\circ$ and $\theta^{}_{13}
\rightarrow 0^\circ$, which guarantee $P^{}_{\tau \alpha} =
P^{}_{\mu \alpha}$ (for $\alpha = e, \mu, \tau$) to hold in the
leading-order approximation. We can then simplify Eq. (12) in the
$\xi \rightarrow 0^\circ$ limit and arrive at the following
result:
\begin{equation}
R^{}_\alpha |^{}_{\xi \rightarrow 0^\circ} \; = \; \frac{P^{}_{\mu
\alpha} + P^{}_{\tau \alpha} \tan^2 \zeta}{\sec^2 \zeta -
\left(P^{}_{\mu \alpha} + P^{}_{\tau \alpha} \tan^2 \zeta\right)}
\; = \; \frac{P^{}_{\mu \alpha}}{1 - P^{}_{\mu \alpha}} \; ,
\end{equation}
where $\zeta$ is completely cancelled out. Therefore, the tiny
dependence of $R^{}_\alpha$ on $\zeta$ appearing in FIG. 4 is just
a natural consequence of the ${\cal O}(\sin\theta^{}_{13})$ or
${\cal O}(\sin^2 \theta^{}_{13})$ corrections to Eq. (16).
\end{enumerate}
Although the numerical examples taken above can only serve for
illustration, they {\it do} give us a ball-park feeling of the
correlation between the source parameters $(\xi, \zeta)$ and the
neutrino mixing parameters $(\theta^{}_{12}, \theta^{}_{23},
\theta^{}_{13}, \delta)$ via the working observables $(R^{}_e,
R^{}_\mu, R^{}_\tau)$ at a neutrino telescope. This observation is
certainly encouraging and remarkable.

\section{Summary}

We have proposed a simple parametrization of the initial flavor
composition of ultrahigh-energy neutrino fluxes generated from
very distant astrophysical sources: $\phi^{}_e : \phi^{}_\mu :
\phi^{}_\tau = \sin^2 \xi \cos^2 \zeta : \cos^2 \xi \cos^2 \zeta :
\sin^2 \zeta$. The conventional mechanism and the postulated
scenarios for cosmic neutrino production can all be reproduced by
taking the special values of $(\xi, \zeta)$. Of course, such a
parametrization is by no means unique. An alternative,
\begin{equation}
\phi^{}_e : \phi^{}_\mu : \phi^{}_\tau \; = \; \frac{x}{1+t} :
\frac{1-x}{1+t} : \frac{t}{1+t} \;
\end{equation}
with $x \in [0, 1]$ and $t \in [0, \infty)$ in general, is also
simple and useful. For a realistic astrophysical source, it should
be more reasonable to take $0 \leq t \ll 1$. Comparing between
Eqs. (4) and (17), one can immediately arrive at $x = \sin^2 \xi$
and $t = \tan^2 \zeta$. The $(x,t)$ and $(\xi, \zeta)$ languages
are therefore equivalent to each other.

After defining three neutrino flux ratios $R^{}_\alpha$ (for
$\alpha = e, \mu, \tau$) as our working observables at a neutrino
telescope, we have shown that the source parameters $\xi$ and
$\zeta$ can in principle be determined by the measurement of two
independent $R^{}_\alpha$ and with the help of accurate neutrino
oscillation data. The standard pion-decay source, the postulated
neutrino beam source and the possible muon-damped source have been
slightly modified to illustrate the sensitivities of $R^{}_\alpha$
to small departures of $\xi$ and $\zeta$ from their given values.
We have also examined the dependence of $R^{}_\alpha$ upon the
smallest neutrino mixing angle $\theta^{}_{13}$ and upon the Dirac
CP-violating phase $\delta$. Our numerical examples indicate that
it is quite promising to determine or constrain the initial flavor
content of ultrahigh-energy neutrino fluxes with the
second-generation neutrino telescopes.

How to measure $R^{}_\alpha$ to an acceptable degree of accuracy
is certainly a big challenge to IceCube and other neutrino
telescopes. A detailed analysis of the feasibility of our idea for
a specific neutrino telescope is desirable, but it is beyond the
scope of this paper. Here we remark that our present understanding
of the production mechanism of cosmic neutrinos needs the
observational and experimental support. We expect that neutrino
telescopes may help us to attain this goal in the long run.

\acknowledgments{One of us (Z.Z.X.) would like to thank X.D. Ji
for his warm hospitality at the University of Maryland, where part
of this work was done. He is also grateful to Z. Cao for useful
discussions, and to R. McKeown and P.D. Serpico for helpful
communications. Our research is supported in part by the National
Natural Science Foundation of China.}

\begin{figure}[t]
\vspace{-2.5cm}
\epsfig{file=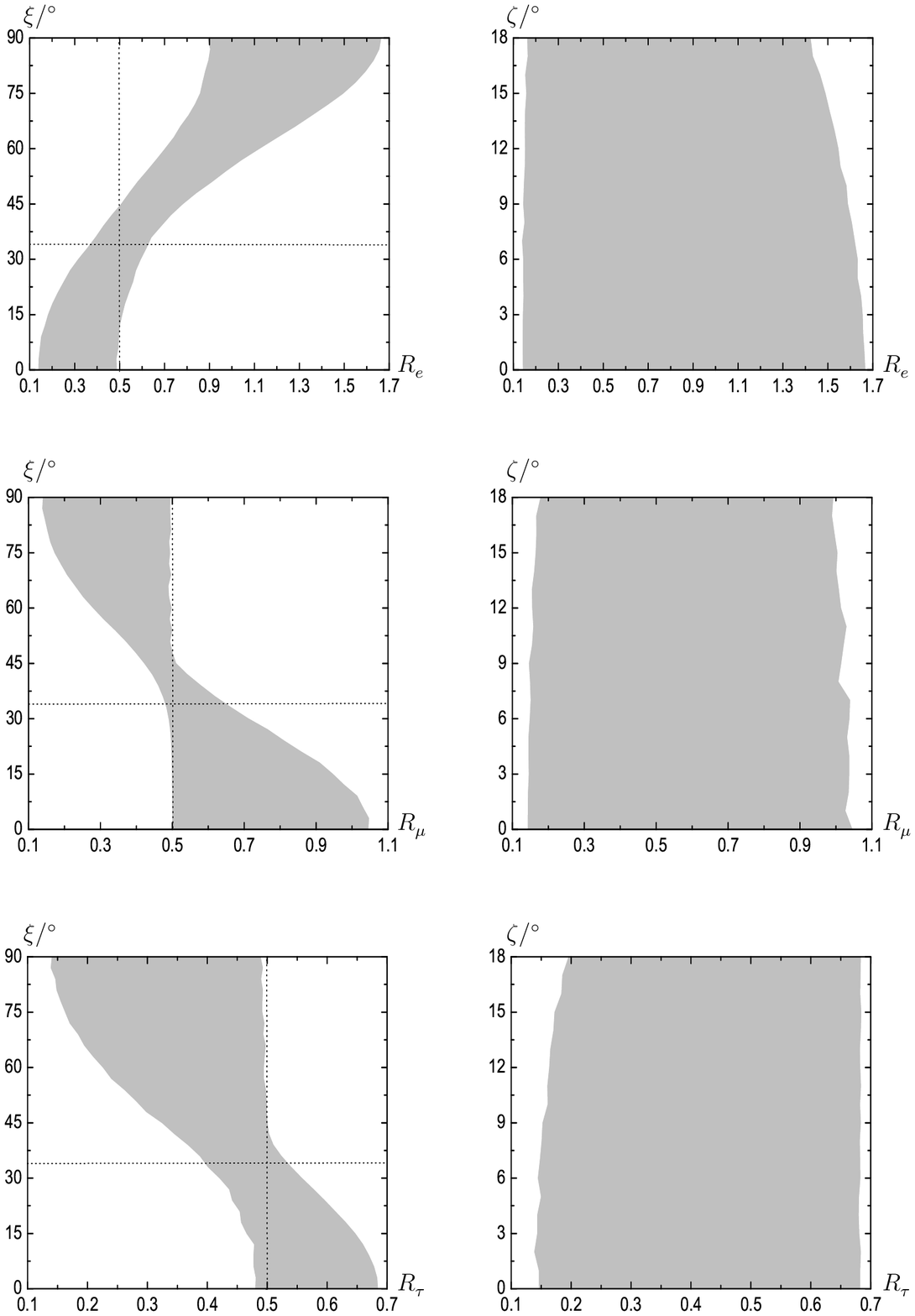,bbllx=2.0cm,bblly=17.5cm,bburx=12cm,bbury=29.5cm,%
width=9cm,height=10cm,angle=0,clip=0}
\vspace{12.5cm}
\caption{Allowed regions of the neutrino flux ratios $R^{}_\alpha$
($\alpha = e, \mu, \tau$) versus the source parameters $\xi$ and
$\zeta$, where we have scanned the $99\%$ C.L. intervals of three
neutrino mixing angles and taken the Dirac CP-violating phase
$\delta \in [0^\circ, 180^\circ]$. The horizontal and vertical
lines in the $(R^{}_\alpha, \xi)$ plots correspond to $\xi =
35.3^\circ$ and $R^{}_\alpha = 0.5$, respectively.}
\end{figure}

\begin{figure}
\epsfig{file=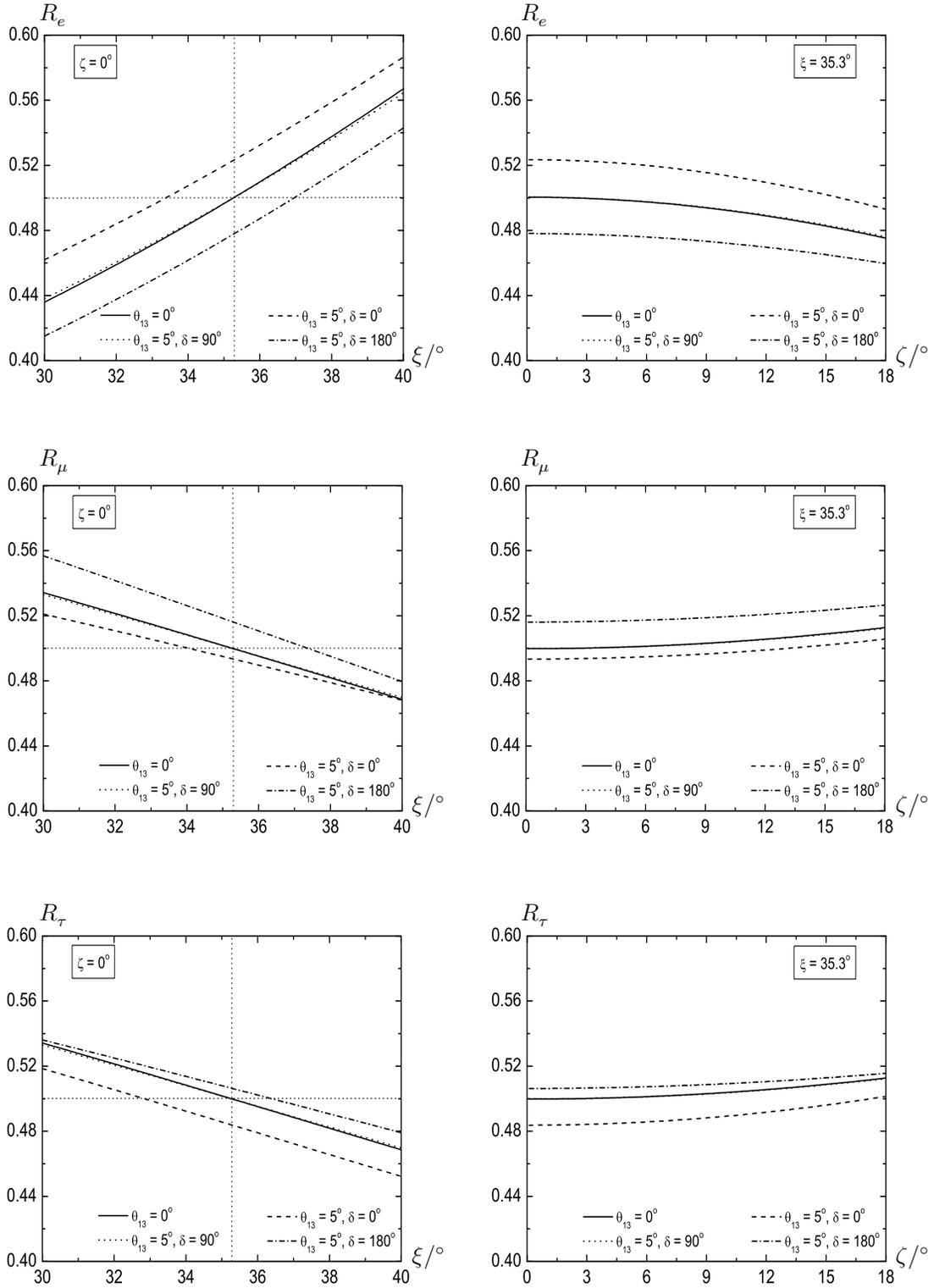,bbllx=2.0cm,bblly=15.2cm,bburx=12cm,bbury=26.5cm,%
width=9cm,height=10cm,angle=0,clip=0}
\vspace{11.cm}
\caption{Numerical illustration of \underline{\bf scenario A},
where $\theta^{}_{12} = 34^\circ$ and $\theta^{}_{23} = 45^\circ$
have typically been input in our calculations. The horizontal and
vertical lines in the $(\xi, R^{}_\alpha)$ plots correspond to
$R^{}_\alpha = 0.5$ and $\xi = 35.3^\circ$, respectively.}
\end{figure}

\begin{figure}
\epsfig{file=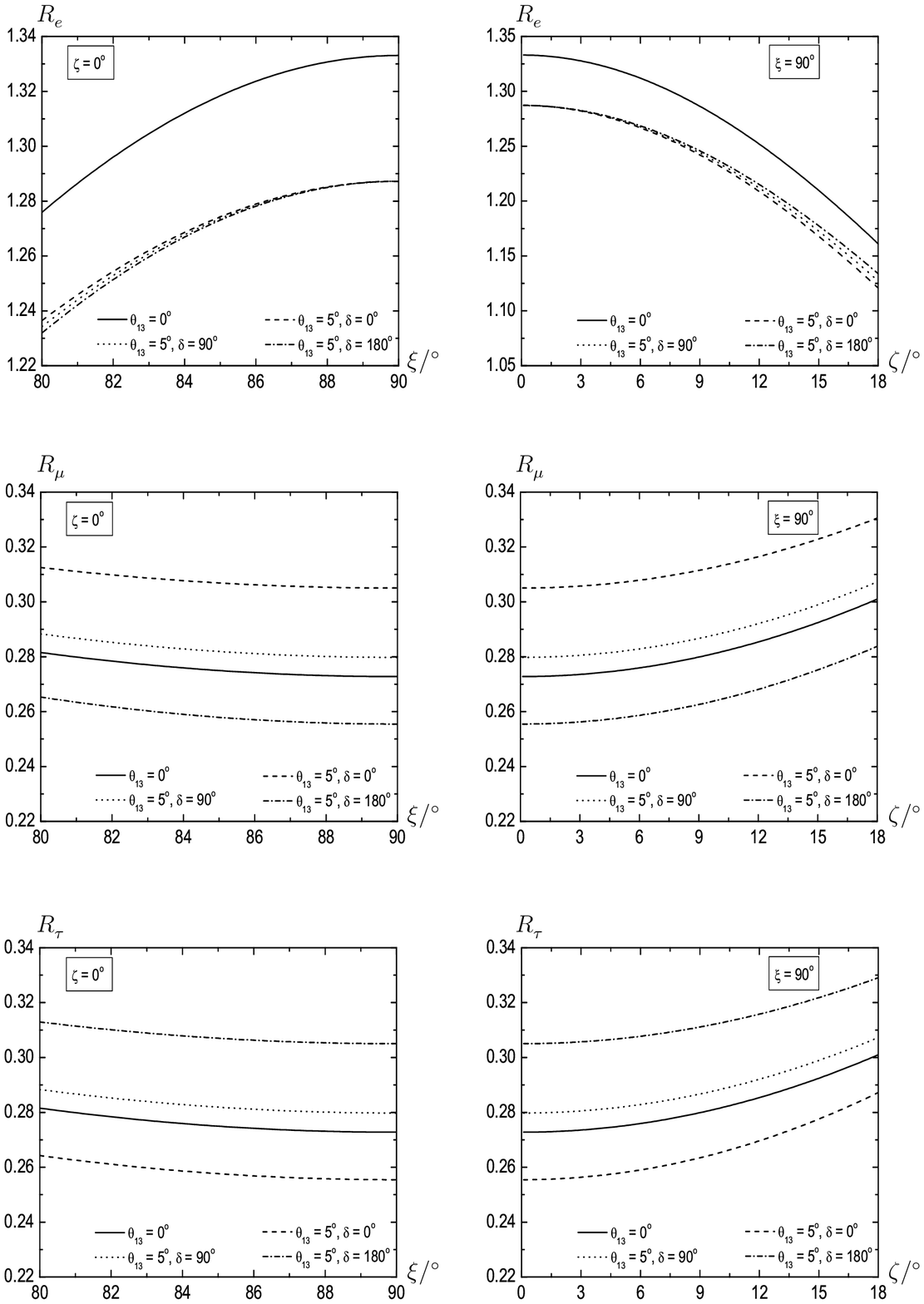,bbllx=2.0cm,bblly=15.2cm,bburx=12cm,bbury=26.5cm,%
width=9cm,height=10cm,angle=0,clip=0}
\vspace{11.cm}
\caption{Numerical illustration of \underline{\bf scenario B},
where $\theta^{}_{12} = 34^\circ$ and $\theta^{}_{23} = 45^\circ$
have typically been input in our calculations.}
\end{figure}

\begin{figure}
\epsfig{file=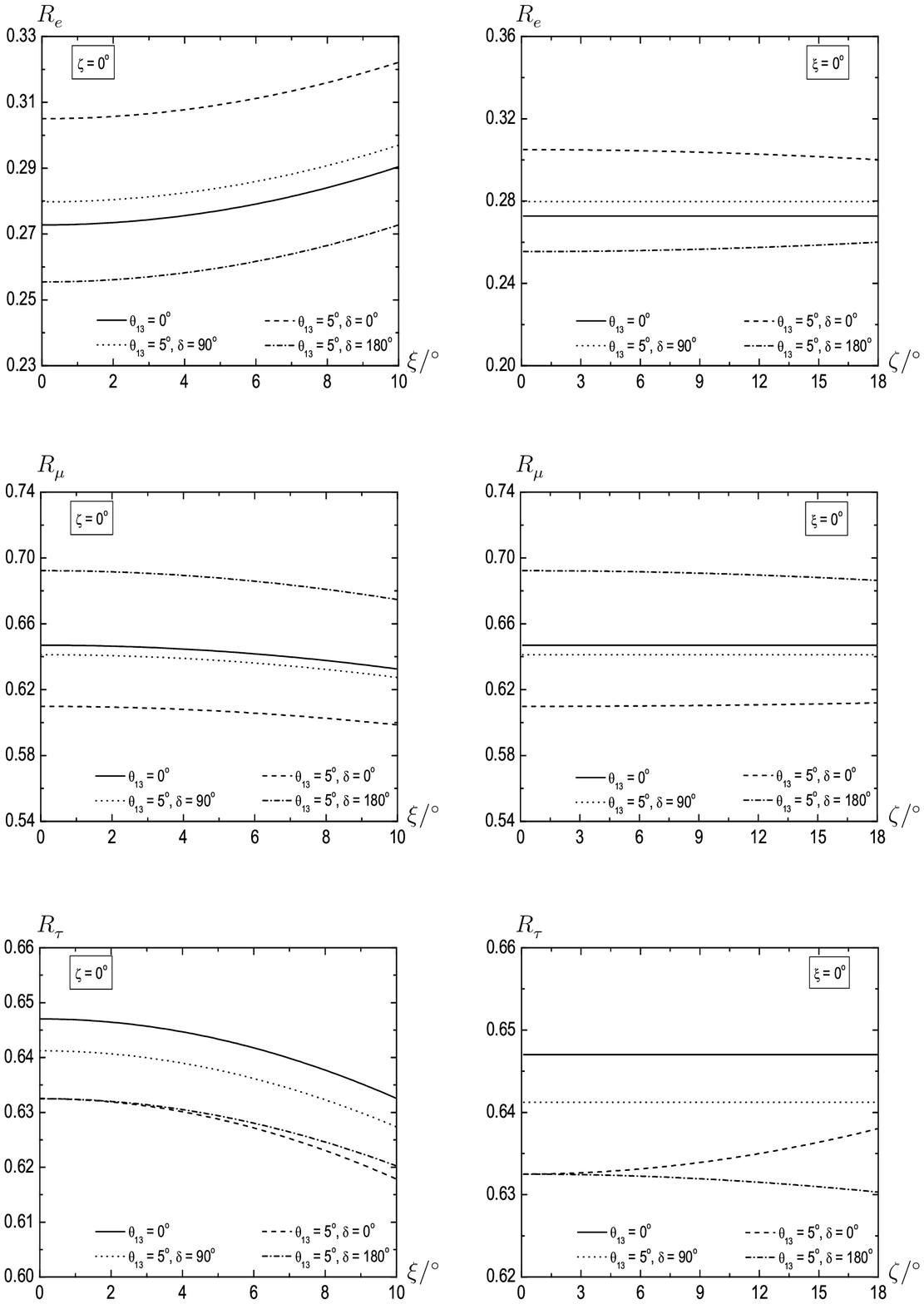,bbllx=2.0cm,bblly=15.2cm,bburx=12cm,bbury=26.5cm,%
width=9cm,height=10cm,angle=0,clip=0}
\vspace{11.cm}
\caption{Numerical illustration of \underline{\bf scenario C},
where $\theta^{}_{12} = 34^\circ$ and $\theta^{}_{23} = 45^\circ$
have typically been input in our calculations.}
\end{figure}

\end{document}